\documentclass[aps,prc,twocolumn,showpacs,superscriptaddress,groupedaddress,a4paper]{revtex4}
\usepackage{graphicx}
\usepackage{epstopdf,color}
\begin{document}
\title{Halo breakup and the Coulomb-nuclear interference problem}

\author{B. Mukeru}
\email{mukerb1@unisa.ac.za}
\affiliation{Department of Physics, University of South Africa, P.O. Box 392, Pretoria 0003, South Africa}

\author{J. Lubian}
\email{lubian@if.uff.br}
\affiliation{Instituto de F\'isica, Universidade Federal Fluminense, Avenida Litoranea s/n, Gragoat\'a, Niter\'oi, RJ, 24210-340, Brazil}

\author{Lauro Tomio}
\email{lauro.tomio@unesp.br}
\affiliation{Instituto de F\'isica T\'eorica, Universidade Estadual Paulista, 01140-070 S\~ao Paulo, SP, Brazil}

\begin{abstract}
  The Coulomb-nuclear interference is studied as a function of the projectile ground-state binding energy  in the 
 $^8{\rm Li}+{}^{12}{\rm C}$ and $^8{\rm Li}+{}^{208}{\rm Pb}$ breakup reactions, by considering an arbitrary range for 
 the $^8{\rm Li}$ ground-state binding energies 
  $\varepsilon_b$, varying from the experimental one 2.03 MeV down to 0.01 MeV. 
 Regardless the target mass, we first show that the Coulomb breakup cross section 
 is stronger dependent on $\varepsilon_b$ than the nuclear breakup cross section, due to the long-range nature of 
the Coulomb forces and to the electromagnetic 
transition matrix elements. For example, in case of $^8{\rm Li}+{}^{208}{\rm Pb}$ reaction at $E_{ lab}=60\,{\rm MeV}$, 
it is found that $|\sigma_{\rm int}|\simeq 4\times \sigma_{\rm nucl}$, while $\sigma_{\rm Coul}\simeq 35\times \sigma_{\rm nucl}$. 
This shows clearly that small nuclear contribution in a Coulomb-dominated reaction does not imply insignificant 
Coulomb-nuclear interference. Such result can be mainly attributed to peripheral interference phenomenon, 
represented by a function of the binding energy, which determines the peripheral range of nuclear forces, 
where Coulomb and nuclear forces strongly interfere destructively.
\end{abstract}


\maketitle

\section{Introduction}
The Coulomb-nuclear interference in halo breakups, has been revealed as an important phenomenon to be
investigated; with considerable efforts been made to understand its role in the breakup process (see, for example, Refs.\,\cite{Hor10,Thomp20,Hussein20,Chat10,2002Margueron,2009Lubian,2009Canto,2006Canto,Kucuk10,Mukeru10,Mukeru20,Tarutina10,Pierre100}). 
However, despite the advances verified by these studies, 
the dynamics around this phenomenon is far from being fully established. 
The question on how both Coulomb and nuclear forces interfere to produce a total breakup remains, to the best of our knowledge,
unsettled. Some of the challenges emanate from the fact that small contribution of the nuclear breakup in a Coulomb-dominated 
reaction does not automatically imply insignificant Coulomb-nuclear interference. This has made the quest for a pure Coulomb 
breakup a formidable task~\cite{Nakamura10,Noc10,Aum10,Fukuda10,Abu10}. Why small nuclear breakup does amount to small 
Coulomb-nuclear interference in a Coulomb-dominated reaction, is an issue that is yet to be addressed. 
Is this also the case in a nuclear-dominated reaction? 
Naturally, a low breakup threshold leads to peripheral collisions, where the Coulomb breakup prevails over the nuclear 
breakup due to mainly the long-range nature of Coulomb forces. 
In this peripheral region (large distance), the Coulomb breakup cross section depends on the projectile structure through 
the electromagnetic matrix element inside the projectile. In fact, for the Coulomb dissociation 
method~\cite{Bertulani10,Winther10,Baur10,Baur20}, the breakup cross section is simply the product of the reaction parameters 
and the projectile electric response function. Therefore, the further the binding energy decreases, the more peripheral the reaction 
becomes, implying in a strong dependence 
on the Coulomb breakup of the projectile
and reaction dynamics. 
Consequently, as the projectile ground-state binding energy decreases, the ratio of the Coulomb breakup cross section over the 
nuclear counterpart is expected to significantly rise regardless the target mass.
In this case, it follows that, one would expect  intuitively that the total breakup cross section should be comparable with the 
Coulomb one, 
due to both dynamic and static effects, which will lead to insignificant Coulomb-nuclear interference. However, a low 
ground-state binding energy is associated with a longer tail of the ground-state wave function. As a result, nuclear forces 
are fairly stretched beyond the projectile-target radius, such that the nuclear breakup cross section is also expected to 
increase as the projectile ground-state binding energy decreases, owing mainly to its dependence of the projectile ground 
state wave function (static effect). How does this play out on the Coulomb-nuclear interference is an interesting question. 
This makes the dependence of the Coulomb-nuclear interference on the projectile ground-state binding energy an important 
study, as it would pave way to elucidate the above-mentioned questions.
\\
\indent The analysis of the dependence of various reaction observables on the projectile ground-state binding energy, is not a novel 
idea (see for example Refs.\,\cite{Wang10,Rath100}, with references therein, and more recently, Ref.\cite{Lei100}). 
However, most studies consider different projectiles of different binding energies.  One of the drawbacks of such an approach, 
is that all the projectiles do not have the same ground-state structure, mass and charge, whose effect on the results in this case 
is not accounted for. Among other ways to circumvent such shortcomings, one can consider different binding energies of the 
same projectile (i.e keeping $N$ and $Z$ unchanged), as done for instance in Refs.\cite{Lei100,Mukeru15}. 
Even though a nucleus would not have different physical ground-state energies, this procedure is more convenient for such 
study. In this paper, we study the breakup of $^8{\rm Li}$ nucleus on the light $^{12}{\rm C}$ and heavy $^{208}{\rm Pb}$ targets. 
The main aim is to investigate the dependence of the Coulomb-nuclear interference on the $^8{\rm Li}$ ground-state binding energy. As stated above, this study will shed light on a better understanding of this interference. We are also interested in investigating in more detail, the fact that in nuclear reactions induced by halo nuclei, an insignificant contribution of the nuclear breakup in a Coulomb-dominated reaction does not amount to insignificant Coulomb-nuclear interference. As a first step, we want to show that, regardless the target mass, as the projectile binding energy decreases, the reaction is dominated by Coulomb breakup. To this end, we consider different arbitrary $^8{\rm Li}$ ground-state binding energies in the range $0.01\,{\rm MeV}\le\varepsilon_b\le 2.033\,{\rm MeV}$, where 2.03\,MeV is the experimental ground-state neutron binding energy\cite{Moro200}. The choice of this projectile is motivated on one hand, by the fact that its breakup has revealed some unusual behavior \cite{Cook20,Pak20,Gum20}.  The total, Coulomb and nuclear breakup cross sections are obtained after a numerical solution of the CDCC (Continuum Discretized Coupled Channels) coupled differential equations \cite{Aust100}, using Fresco codes \cite{Thompson100}.\\
\indent The paper is organized as follows: In Section \ref{calculation}, the details of the calculations are presented together with a brief summary of the CDCC formalism. The results are presented and discussed in Section \ref{results}, while the conclusions are summarized in Section \ref{conclusion}.

\section{Calculations Details}
\label{calculation}
\subsection{Brief description of CDCC formalism}

In the CDCC formalism, the coupled differential equations are of the following form
\begin{eqnarray}\label{coupled}
\big[{T}_R&+&U_{\alpha\alpha}^{LJ}(R)+\varepsilon_{\alpha}
-E \big]\chi_{\alpha}^{LJ}(R)\nonumber\\
&-&\sum_{\alpha\ne\alpha'}i^{L-L'}U_{\alpha\alpha'}^{LL'J}(R)\chi_{\alpha'}^{L'J}(R)=0,
\end{eqnarray}
where 
\begin{eqnarray}
{T}_R=-\frac{\hbar^2}{2\mu_{pt}}\bigg(\frac{d^2}{dR^2}-\frac{L(L+1)}{R^2}\bigg),
\end{eqnarray}
and $U_{\alpha\alpha'}^{LL'J}(R)$ are the potential matrix elements, given by
\begin{eqnarray}\label{potmatr}
U_{\alpha\alpha'}^{LL'J}(R)=\langle\mathcal{Y}_{\alpha L}({\bf r},\Omega_R)|V_{pt}({\bf r, R})|\mathcal{Y}_{\alpha' L'}({\bf r},\Omega_R)\rangle,
\end{eqnarray}
where $\mathcal{Y}_{\alpha L}({\bf r},\Omega_R)$ is a channel wave function defined as 
\begin{eqnarray}\label{channel}
  \mathcal{Y}_{\alpha L}({\bf r},\Omega_R)=[\hat\Phi_{\alpha}({\bf r})\otimes Y_L^{\Lambda}(\Omega_R)]_{JM},
\end{eqnarray}
with $\hat\Phi_{\alpha}({\bf r})$ the wave function containing the square integrable discretized bin wave functions. The subscript [$\alpha=(i,\ell,s,j), i=0,1,2,\ldots,N_b$, $N_b$ being the number of bins] represents all the relevant quantum numbers that describe different states of the projectile, where the ground-state refers to $i=0$. Given that the projectile-target optical potential is $V_{pt}({\bf r, R})=V_{ct}({\bf R}_{ct})+V_{nt}({\bf R}_{nt})$, $V_{ct}({\bf R}_{ct})$ and $V_{nt}({\bf R}_{nt})$ the core ($^7{\rm Li}$)-target and neutron-target optical potentials, respectively, where $ {\bf R}_{ct}={\bf R}+\frac{1}{8}{\bf r}, {\bf R}_{nt}={\bf R}-\frac{7}{8}{\bf r}$, with the vector ${\bf R}$ being the projectile-target centre of mass coordinate, and ${\bf r}$ the internal coordinate of the two-body projectile, the potential matrix elements (\ref{potmatr}) can be decomposed as follows
 \begin{eqnarray}\label{potcomp}
   U_{\alpha\alpha'}^{LL'J}(R)&=& U_{\alpha\alpha'}^{Coul}({\bf r,R})+ U_{\alpha\alpha'}^{Nucl}({\bf r,R})\nonumber\\
&=&\langle\mathcal{Y}_{\alpha L}({\bf r},\Omega_R)|V_{ct}({\bf R}_{ct})|\mathcal{Y}_{\alpha' L'}({\bf r},\Omega_R)\rangle\nonumber\\
  &+&\langle\mathcal{Y}_{\alpha L}({\bf r},\Omega_R)|V_{nt}({\bf R}_{nt})|\mathcal{Y}_{\alpha' L'}({\bf r},\Omega_R)\rangle\\
  &=&\langle\mathcal{Y}_{\alpha L}({\bf r},\Omega_R)|V_{ct}^{Coul}({\bf R}_{ct})|\mathcal{Y}_{\alpha' L'}({\bf r},\Omega_R)\rangle\nonumber\\
  &+&\langle\mathcal{Y}_{\alpha L}({\bf r},\Omega_R)|V_{ct}^{nucl}({\bf R}_{ct})|\mathcal{Y}_{\alpha' L'}({\bf r},\Omega_R)\rangle\nonumber\\
  &+&\langle\mathcal{Y}_{\alpha L}({\bf r},\Omega_R)|V_{nt}^{nucl}({\bf R}_{nt})|\mathcal{Y}_{\alpha' L'}({\bf r},\Omega_R)\rangle\nonumber,
\end{eqnarray}
\noindent since $V_{nt}^{Coul}({\bf R}_{nt})=0$, for a neutron-halo projectile.
Therefore, Coulomb breakup corresponds to $U_{\alpha\alpha'}^{LL'J}(R)\to U_{\alpha\alpha'}^{Coul}(R)$, and the nuclear breakup to $U_{\alpha\alpha'}^{LL'J}(R)\to U_{\alpha\alpha'}^{nucl}(R)$. In other words, the Coulomb breakup is obtained when nuclear forces are switched off and vice versa. While there are some ambiguities in separating the total breakup cross section into its Coulomb and nuclear components as reported in Ref.\cite{Pierre100}, the procedure adopted in this paper is fairly adequate for the problem at hand. After a numerical evaluation of the coupling matrix elements (\ref{potmatr}), the coupled differential equations (\ref{coupled}) are solved with the usual boundary conditions at large distance
\begin{eqnarray}\label{BC}
  \chi_{\alpha}^{LJ}(R)\stackrel{R\to\infty}\longrightarrow \frac{ i}{2}\left[H_{\alpha}^-(K_{\alpha}R)\delta_{\alpha\alpha'}-H_{\alpha}^+(K_{\alpha}R)S_{\alpha\alpha'}^{LL'J}\right],
\end{eqnarray}
where $H_{\alpha}^{\pm}(K_{\alpha}R)$ are incoming and outgoing Coulomb Hankel functions \cite{Abramo100}, and $S_{\alpha\alpha'}(K_{\alpha})$ is the breakup S-matrix, with the wave number $K_{\alpha}=\sqrt{\frac{2\mu_{pt}(E+\varepsilon_{\alpha})}{\hbar^2}}$. Notice that the potential matrix elements $ U_{\alpha\alpha'}^{nucl}(R)$ vanished at large distance:
{\small\begin{eqnarray}\label{potnucl}
  U_{\alpha\alpha'}^{nucl}(R)\stackrel{R> R_n}\longrightarrow 0,\;{\rm with}\;R_n\equiv r_0(A_p^{1/3}+A_t^{1/3})+\delta_R(\varepsilon_b).
  \end{eqnarray}}
The function $\delta_R(\varepsilon_b)$ is introduced to account for the fact that, for low breakup thresholds, nuclear forces are slightly stretched beyond $R_0=r_0(A_p^{1/3}+A_t^{1/3})$. Therefore, in this case, the coupled differential equations (\ref{coupled}), are reduced to uncoupled differential equations whose solution give the nuclear S-matrix. The various breakup cross sections are obtained using the relevant S-matrix as outlined for example in Ref. \cite{Thompson100}.

As mentioned in the introduction, at large distance ($R\to\infty$), the Coulomb breakup directly depends on the projectile response function. To explicitly show this, we first expand the projectile-target Coulomb interaction into potential multipoles as follows \cite{Hussein50}
{\small
\begin{eqnarray}\label{CE}
  V^{Coul}({\bf r},{\bf R})\stackrel{R\to\infty}\longrightarrow 4\pi Z_te\sum_{\lambda=0}^{\lambda_{\rm max}}\frac{\hat \lambda}{R^{\lambda+1}}\left[\mathcal{M}_{\lambda}^{\varepsilon}({\bf r})\otimes Y_{\lambda}(\Omega_R)\right]_{\lambda 0},
  \end{eqnarray}}
where $\lambda$ is the multipole order, truncated by $\lambda_{\rm max}$, $\hat\lambda=(2\lambda+1)^{1/2}$, and $\mathcal{M}_{\lambda}^{\varepsilon}({\bf r})$ the electric operator of the projectile given by
{\small 
\begin{eqnarray}\label{EO}
  \mathcal{M}_{\lambda}^{\varepsilon}({\bf r})&=&e\left[Z_n\left(\frac{A_c}{A_p}\right)^{\lambda}+Z_c\left(-\frac{A_n}{A_p}\right)^{\lambda}\right]r^{\lambda}Y_{\lambda}^{\mu}(\Omega_r)\nonumber\\
          [2mm]
         &=&e\left[Z_c\left(-\frac{A_n}{A_p}\right)^{\lambda}\right]r^{\lambda}Y_{\lambda}^{\mu}(\Omega_r),
  \end{eqnarray}}
since $Z_n=0$, in our case. The projectile response function for the transition from the projectile ground-state to the continuum states can be obtained through $\mathcal{M}_{\lambda}^{\varepsilon}({\bf r})$ as follows \cite{Bertulani50}
{\small
  \begin{eqnarray}\label{electric}
 \frac{dB(E\lambda)}{d\varepsilon}&=&\frac{\mu_{cn}}{\hbar^2 k}\sum_j(2j+1)|\langle \Phi_{n\ell_0}^{j_0}({\bf r})|\mathcal{M}_{\lambda}^{\varepsilon}({\bf r})|\Phi_{k\ell}^{j}({\bf r})|^2\nonumber\\
       [2mm]
       &=&\frac{\mu_{cn}}{\hbar^2 k}\sum_j(2j+1)|\mathcal{F}(r)|^2,
\end{eqnarray}}
where $\mu_{cn}$ is the core-neutron reduced mass, and
{\small \begin{eqnarray}
 \mathcal{F}(r)&=& eZ_{\lambda}(-1)^{\ell_0+\ell+j+s+\lambda}\left(\begin{array}{ccccc}
    \ell_0 & \lambda & \ell \\
    0 & 0& 0 
 \end{array}
 \right)
 \left \{\ \begin{array}{cccc}
 s &\ell_0 & j_0 \\
 \lambda & j & \ell
\end{array}
 \right\}\
 \nonumber\\
 & \times & \int_0^{\infty} \phi_{n\ell_0}^{j_0}(r)r^{\lambda}\phi_{k\ell}^j(r) dr, 
\end{eqnarray}}
with $\phi_{n\ell_0}^{j_0}(r)$, and $\phi_{k\ell}^j(r)$, the ground-state and continuum radial wave functions. Eqs.(\ref{CE}), and (\ref{electric}), explicitly indicate how the Coulomb breakup depends on the projectile structure. For example, following \cite{Bertulani10,Winther10,Baur10,Baur20}, and using the recurrence relation \cite{Abramo100}, it can be shown that the first-order energy distributions Coulomb breakup cross section is given in terms of Eq.(\ref{electric}), by
{\small \begin{eqnarray}\label{break1}
\frac{d\sigma^{\rm Coul}}{d\varepsilon}=\frac{32\pi^2}{9}\bigg(\frac{Z_Te}{\hbar
  v}\bigg)^2z_{\rm min} K_0(z_{\rm min})K_1(z_{\rm min})\frac{dB(E1)}{d\varepsilon},
\end{eqnarray}}
where $K_x$ are modified Bessel functions \cite{Abramo100}, and 
\begin{eqnarray}\label{zmin}
  z_{ \rm min}=\frac{\omega}{v}b_{ \rm min}=\frac{\varepsilon-\varepsilon_b}{\hbar v}b_{\rm min},
\end{eqnarray}
with 
\begin{eqnarray}\label{impar}
b_{ \rm min}=\frac{Z_pZ_te^2}{2E\tan(\theta_c/2)},
\end{eqnarray}
is the minimum impact parameter, where $\theta_c$ is defined as the maximum cutoff scattering angle up to which the Coulomb breakup is dominant.

\subsection{Conditions of calculations}

To construct the set of coupled differential equations (\ref{coupled}), one needs the bound and continuum wave functions of the 
${}^7{\rm Li}+n$ system. The ground-state binding energy is $\varepsilon_b=2.033\,{\rm MeV}$, with $j^{\pi}=2^+$. The system 
also exhibits a bound excited state of $\varepsilon_{\rm ex}=0.88\,{\rm MeV}$, and $j^{\pi}=1^+$ \cite{Moro200}. These wave 
functions are obtained after a numerical solution of the two-body Schr\"odinger equation with a Woods-Saxon potential as input 
whose parameters are $V_{\rm SO}=4.89\,{\rm MeV\cdot fm^2}$, $r_0=r_{\rm SO}=1.25\,{\rm fm}$ and $a_0=a_{\rm SO}=0.52\,{\rm fm}$, 
taken from \cite{Moro200}. The depth $V_0$ of the central part was adjusted to reproduce the ground and bound excited state energies. 
Similarly, the other binding energies are obtained by adjusting $V_0$.
\begin{figure}[t]
\begin{center}
  \resizebox{82mm}{!}{\includegraphics{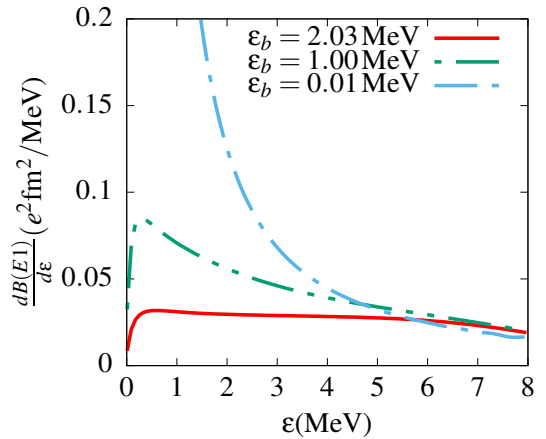}}
\end{center}
\caption{\label{fig01}Electric response function $B(E1)$ for the $^7{\rm Li}+n$ system, for the transition from to continuum 
$s$- plus $d$-states.The plotted behaviors are shown for three different $^8{\rm Li}$ ground-state binding energies, as 
indicated.
}
\end{figure}
\begin{figure}[t]
\begin{center}
  \resizebox{82mm}{!}{\includegraphics{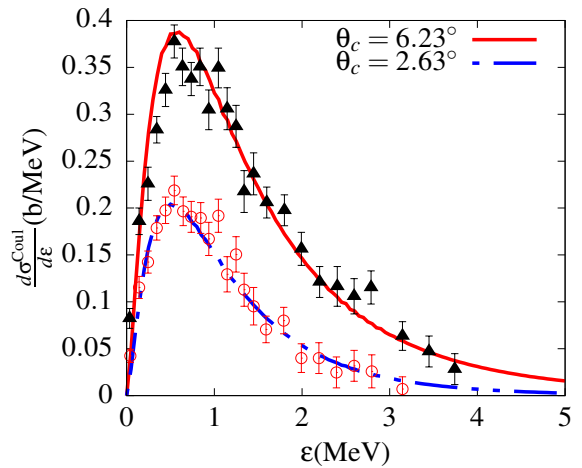}}
\end{center}
\caption{\label{fig02}Coulomb breakup model results, obtained for $^{15}$C on $^{208}$Pb at 68 MeV/nucleon $(v=0.36c)$ are shown
as function of the relative $^{14}$C$-n$ energy ${\varepsilon}$, for $\theta_c=6.23^\circ$ and 2.63$^\circ$ (as indicated), 
respectively referred to $b_{\rm min}=$7.44 fm and 16.20 fm. 
The corresponding data are from Ref.~\cite{Nakamura350}.}
\end{figure}
With these parameters, we first calculate the electric response function $B(E1)$, given by Eq.(\ref{electric}), 
for different binding energies to display the dependence of the Coulomb breakup on the projectile internal structure. 
To this end, $B(E1)$ is plotted in Fig.\ref{fig01}, for transition from the ground-state to continuum $s$- plus $d$-states, 
for $\varepsilon_b=0.01\,{\rm MeV}, 1.0\,{\rm MeV}$ and $2.03\,{\rm MeV}$. One notices that, $B(E1)$ is substantially 
larger for $\varepsilon_b=0.01\,{\rm MeV}$ compared to its value for $\varepsilon_b=2.03\,{\rm MeV}$ for 
$\varepsilon_b\le 1.5\,{\rm MeV}$. It is therefore clear that in the asymptotic region, the Coulomb breakup has a 
strong dependence on the projectile internal structure. To further show this more clearly, we present in Fig.\ref{fig02}, 
the energy distributions Coulomb breakup cross section for $^{15}{\rm C}+{}^{208}{\rm Pb}$ reaction, as function of the 
$^{15}{\rm C}\to {}^{14}{\rm C}+n$ relative energy {$\large \varepsilon$}. The parameters of the $^{14}{\rm C}+n$ Woods-Saxon potential are 
similar to those of Ref.\cite{Mukeru400}. An excellent fit of the experimental data is observed in this figure.
\\
\indent To evaluate the coupling matrix element (\ref{coupled}), one needs the projectile-target optical potentials. 
For $^{12}{\rm C}$ target, the $^7{\rm Li}+{}^{12}{\rm C}$ and $n+{}^{12}{\rm C}$ optical potential parameters were 
taken from Ref.\,\cite{Barioni100}. The $^7{\rm Li}+{}^{208}{\rm Pb}$ optical potential parameters were obtained from 
the $^7{\rm Li}$ global potential of Ref.\cite{Cook300}, with the depth of the real part slightly modified to fit the elastic 
scattering experimental data. The parameters of the $n+{}^{208}{\rm Pb}$ optical potential were taken 
from \cite{Mukeru300}. The various integration parameters are listed in Table\,\ref{table1}. The adopted bin 
widths were, $\Delta\varepsilon=0.5\,{\rm MeV}$, for $s$- and $p$-states, $\Delta\varepsilon=1.0\,{\rm MeV}$, 
for $f$- and $d$-states and $\Delta\varepsilon=1.5\,{\rm MeV}$ for ${\rm g}$-states. These parameters were 
selected in accordance with the convergence requirements, where $\ell_{\rm max}$, $\lambda_{\rm max}$, 
$r_{\rm max}$, $R_{\rm max}$ and $L_{\rm max}$ are the maximum angular momentum between $^7{\rm Li}$ 
and the neutron, maximum order of the multipole expansion, maximum matching radius for bin integration, maximum 
matching radius of the integration of the coupled differential equations, and the maximum angular momentum of the 
relative centre of mass motion, respectively.

\begin{figure}[t]
\begin{center}
  \resizebox{82mm}{!}{\includegraphics{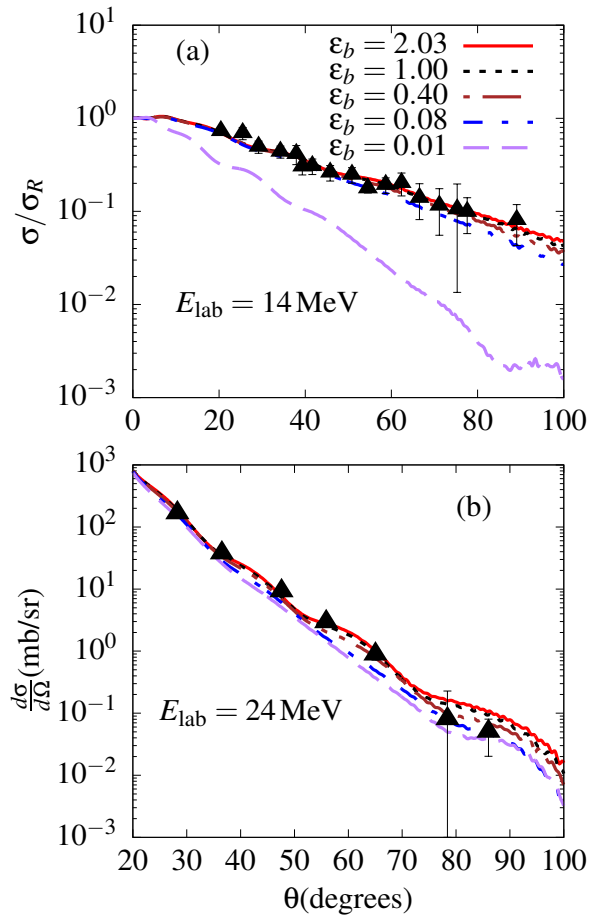}}
\end{center}
\caption{\label{fig03}$^8{\rm Li}+{}^{12}{\rm C}$ elastic scattering cross sections are shown for two incident energies. 
In the upper panel (a), for $E_{lab}=$14 MeV,  $\sigma$ is given in units of the Rutherford cross section $\sigma_R$; 
In the lower panel (b), for $E_{lab}=$24 MeV,  it is shown the differential elastic scattering cross section.
The results in both panels are for different $^8$Li binding energies, as indicated in the upper panel.
The data points are from Ref.~\cite{Jinr20}.}
\end{figure}
\begin{figure}[h]
\begin{center}
  \resizebox{82mm}{!}{\includegraphics{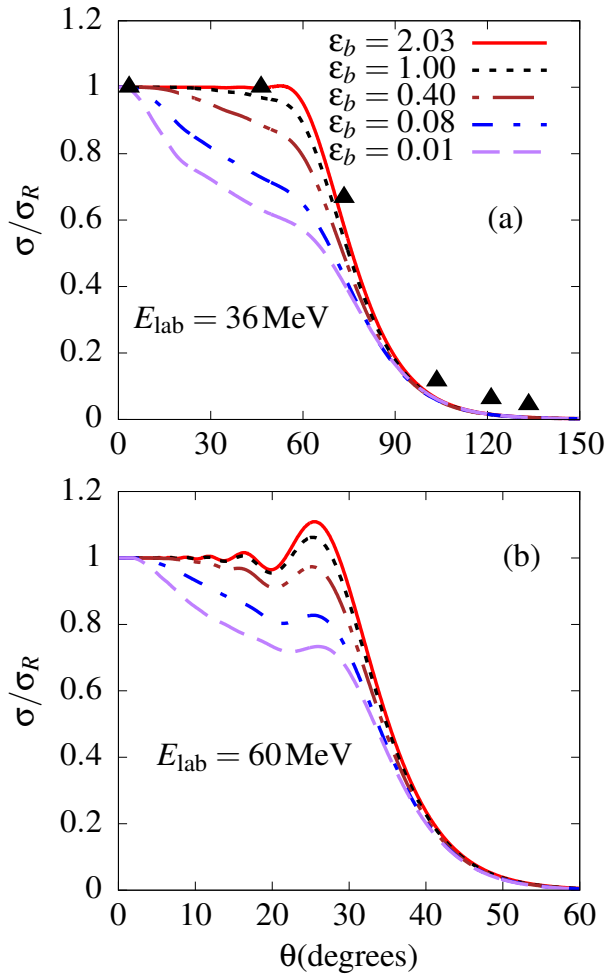}}
\end{center}
\caption{\label{fig04}$^8{\rm Li}+{}^{208}{\rm Pb}$ elastic scattering cross sections, in units of $\sigma_R$, for different 
incident and binding energies. The data points are from Ref.~\cite{Jinr20}.}
\end{figure}

\begin{table}[h]
\caption{\label{table1} CDCC model space parameters required for convergence.}
\begin{tabular}{lllllllll}
\hline\hline
& \multicolumn{1}{c}{$\ell_{\mbox{\tiny{max}}}$} & \multicolumn{1}{c}{$\lambda_{\mbox{\tiny{max}}}$} & \multicolumn{1}{c}{$\varepsilon_{\mbox{\tiny{max}}}$} & \multicolumn{1}{c}{$r_{\mbox{\tiny{max}}}$} & \multicolumn{1}{c}{$L_{\mbox{\tiny{max}}}$} & \multicolumn{1}{c}{$R_{\mbox{\tiny{max}}}$} &\multicolumn{1}{c}{ $\Delta R$}\\
& \multicolumn{1}{c}{} &\multicolumn{1}{c}{} &\multicolumn{1}{c}{(MeV)} & \multicolumn{1}{c}{(fm)} & \multicolumn{1}{c}{} & \multicolumn{1}{c}{(fm)} & \multicolumn{1}{c}{(fm)}\\
\hline
$^{12}{\rm C}:$&\multicolumn{1}{c}{3} &\multicolumn{1}{c}{ 3} & \multicolumn{1}{c}{6} & \multicolumn{1}{c}{80}  & \multicolumn{1}{c}{300} & \multicolumn{1}{c}{300} & \multicolumn{1}{c}{0.08}\\

$^{208}{\rm Pb}:$&\multicolumn{1}{c}{4} &\multicolumn{1}{c}{ 4} & \multicolumn{1}{c}{10} & \multicolumn{1}{c}{80}  & \multicolumn{1}{c}{1000} & \multicolumn{1}{c}{600} & \multicolumn{1}{c}{0.03}\\

\hline\hline
\end{tabular}
\end{table}

\section{Results and Discussion}
\label{results}
\subsection{Elastic scattering cross sections}
\label{elastic}
In this section, we first analyze the dependence of the elastic scattering cross sections on the projectile ground-state binding energy in 
both Coulomb- and nuclear-dominated reactions. These cross sections are plotted in two panels given in Fig.~\ref{fig03}, for the 
$^{12}{\rm C}$ target; and in two panels of Fig.~\ref{fig04}, for $^{208}{\rm Pb}$ target, considering two different incident energies,
with the same set of $^8{\rm Li}$ ground-state binding energies. 
In case of $^{12}{\rm C}$ target [shown in Fig.~\ref{fig03} for $E_{lab}=$14 MeV (a) and 24 MeV (b)], 
which is a nuclear-dominated reaction, one observes a weak dependence on the ground-state binding energy in the range 
$0.4\,{\rm MeV}\le\varepsilon_b\le 2.03\,{\rm MeV}$. 
However, as the binding energy decreases further ($\varepsilon_b\le 0.08\,{\rm MeV}$), a clear decrease of the elastic 
scattering cross section is noticed, which becomes more pronounced for $\varepsilon_b=0.01\,{\rm MeV}$ [as seen in panel (a) 
of Fig.\ref{fig03}].
\begin{figure}[t]
\begin{center}
  \resizebox{85mm}{!}{\includegraphics{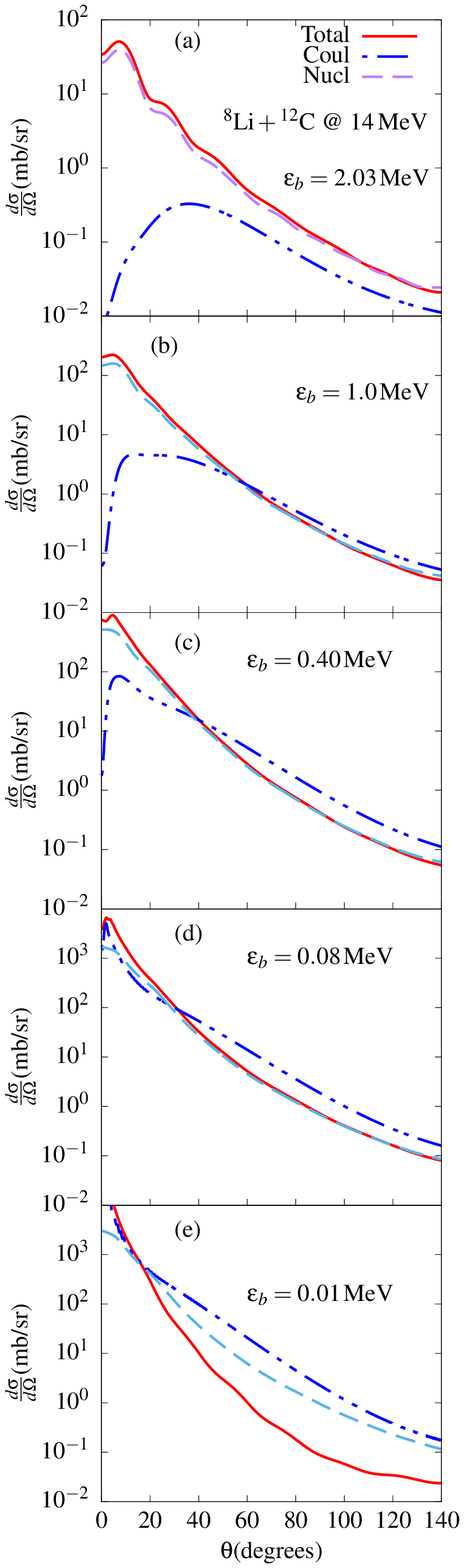}\includegraphics{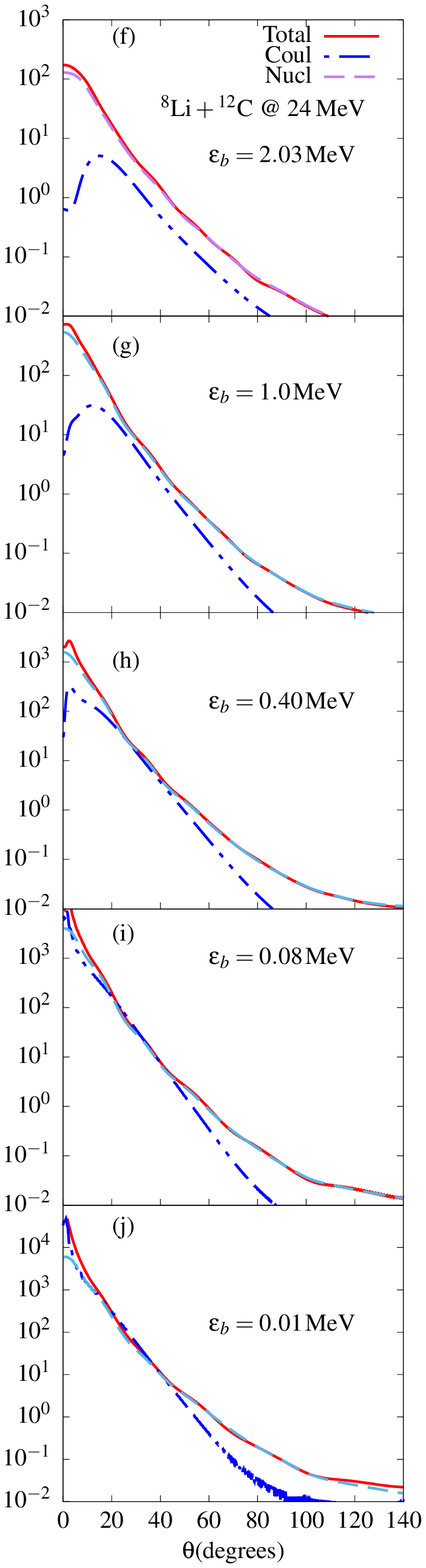}}
\end{center}
\caption{\label{fig05}$^8{\rm Li}+{}^{12}{\rm C}$ angular distributions differential total, Coulomb and nuclear breakup cross sections for different binding energies at $E_{ lab}=14\,{\rm MeV}$ (left panels) and $E_{ lab}=24\,{\rm MeV}$ (right panels).}
\end{figure}
\begin{figure}[!!t]
\begin{center}
  \resizebox{85mm}{!}{\includegraphics{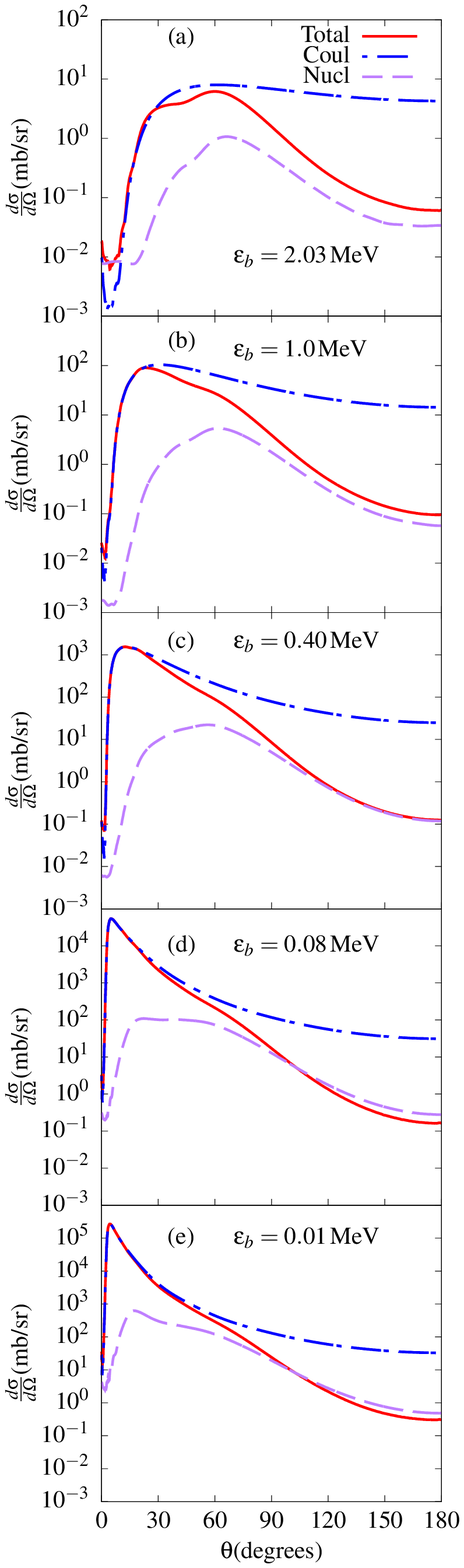}\includegraphics{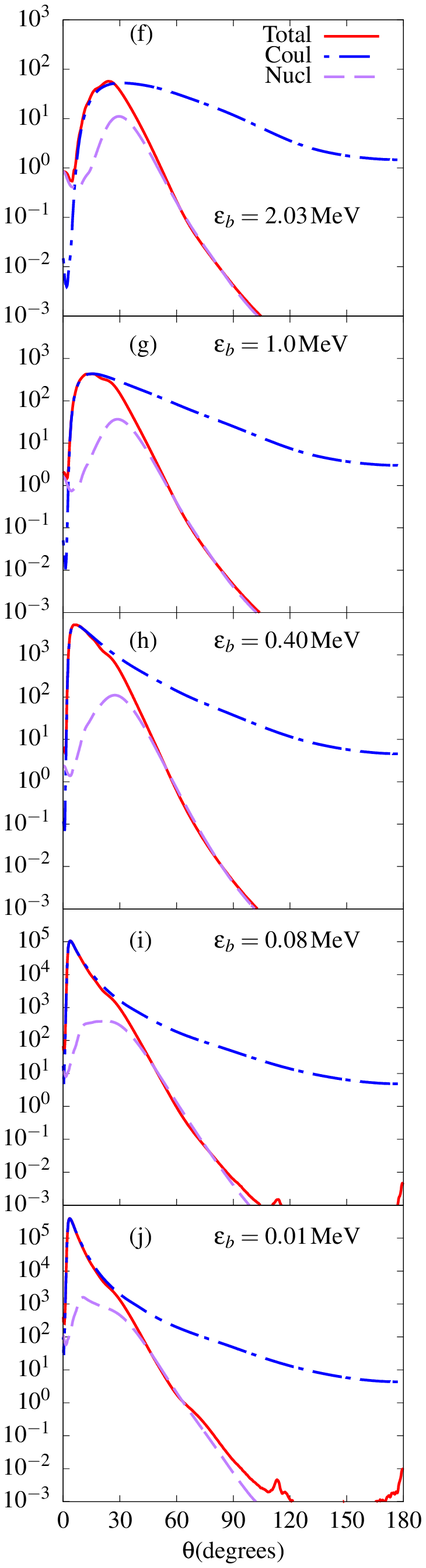}}
\end{center}
\caption{\label{fig06}$^8{\rm Li}+{}^{208}{\rm Pb}$ angular distributions differential total, Coulomb and nuclear breakup cross 
sections, for different binding energies at $E_{ lab}=36\,{\rm MeV}$ (left panels) and $E_{ lab}=60\,{\rm MeV}$ (right panels).}
\end{figure} 
On the other hand, for $^{208}{\rm Pb}$ target [shown in Fig.\,\ref{fig04} for $E_{lab}=$36 MeV (a) and 60 MeV (b)], which is a 
Coulomb-dominated reaction, it is observed a strong dependence of the elastic scattering cross section on all the different binding 
energies at forward angles.  These results reveal that, where the nuclear breakup is dominant, the effect of the binding energy on the elastic 
scattering cross section is relatively small, whereas this effect becomes more pronounced when the Coulomb breakup is dominant. 
Then, as it is observed that an important effect happens for the $^8{\rm Li}+{}^{12}{\rm C}$ reaction when 
$\varepsilon_b\le 0.08\,{\rm MeV}$,  
which is similar to the $^8{\rm Li}+{}^{208}{\rm Pb}$ reaction case, one can imply that the 
$^8{\rm Li}+{}^{12}{\rm C}$ reaction is already dominated by the Coulomb breakup when reaching this binding energy range.
As the binding energy decreases, the Coulomb breakup becomes dominant over its nuclear counterpart, as anticipated. 
It also follows that
the probability of the projectile to fly on the outgoing unbroken trajectory decreases, diminishing the corresponding elastic 
scattering cross section.
We will look into this conclusion in more detail in the next section. 

\subsection{Breakup cross sections}
\label{breakup}
The differential total, Coulomb and nuclear breakup cross sections for $^{12}{\rm C}$ target are depicted in Fig.~\ref{fig05}, 
for a set of projectile binding energies $\varepsilon_b$ from 0.01 MeV up to 2.03 MeV, considering the target energies 
$E_{lab}=$14 MeV (left set of panels) and 24 MeV (right set of panels) energies.
As it should be expected, for $\varepsilon_b=2.03\,{\rm MeV}$, the nuclear breakup cross section is more important than the 
Coulomb breakup cross section and is relatively similar to the total breakup cross section. However, as the binding energy decreases, 
the Coulomb breakup becomes progressively more important. In fact, it is noticed that, for $\varepsilon_b=1.0\,{\rm MeV}$ 
[Fig.~\ref{fig05}(b)], the Coulomb breakup cross section is already slightly important than both total and nuclear breakup cross sections 
at $\theta\ge 60^{\circ}$. Considering $E_{ lab}=24\,{\rm MeV}$ (right panels of Fig.\ref{fig05}), it is seen that even though the nuclear 
breakup cross section remains dominant at backward angles for all binding energies, while the Coulomb breakup appears to prevail 
at forward angles for $\varepsilon_b\le 0.08\,{\rm MeV}$ [see panels \ref{fig05}(i) and \ref{fig05}(j)]. We can therefore conclude that, 
as the binding energy decreases the Coulomb breakup becomes dominant even in a naturally nuclear-dominated reaction, owing to 
the long-range of Coulomb forces on one hand, and to its direct dependence on the electromagnetic transition matrix elements on 
the other hand. This confirms our assessment in section \ref{elastic}. Yet another observation is the competition between the total 
and nuclear breakup cross sections in the binding energy range of $0.08\,{\rm MeV}\le\varepsilon_b\le 2.03\,{\rm MeV}$ 
[Fig.~\ref{fig05}(a)-(d) and Fig.~\ref{fig05}(f)-(j)]. 
By further decreasing $\varepsilon_b$ to 0.01 MeV) [Fig.\ref{fig05}(e)], it is shown that both Coulomb and nuclear breakup 
cross sections become more important than their total breakup counterpart at $\theta\ge 20^{\circ}$. 
This already provide a glimpse on the Coulomb-nuclear interference dependence on the binding energy, which will be 
discussed in section \ref{interf}.\\
\begin{figure}[t]
\begin{center}
  \resizebox{82mm}{!}{\includegraphics{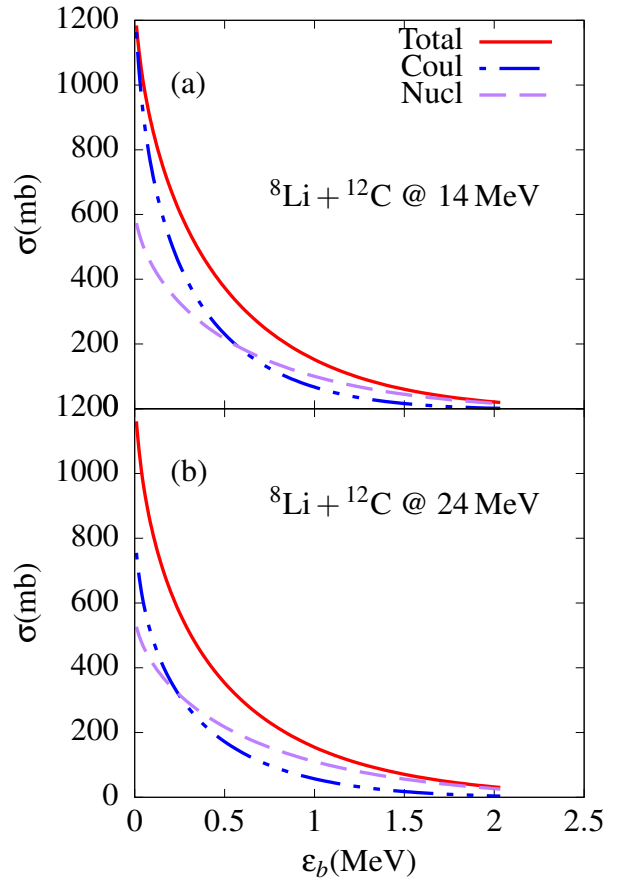}}
\end{center}
\caption{\label{fig07}$^8{\rm Li}+{}^{12}{\rm C}$ angular-integrated total, Coulomb and nuclear breakup cross sections as 
functions of the binding energy.}
\end{figure}
\indent Turning to the $^{208}{\rm Pb}$ target (Fig.~\ref{fig06}), an expected large importance of the Coulomb breakup cross sections 
over total (at backward angles) and nuclear breakup cross section is observed, for all binding and incident energies considered. 
A careful inspection of the panels in this figure indicates that, for $\varepsilon_b\le 1\,{\rm MeV}$, the total and Coulomb breakup 
cross sections are about one order of magnitude larger than the nuclear breakup cross section  [See panels (a) and (b) at 
$\theta\le 60^{\circ}$] and [Fig.\ref{fig06}(f) and \ref{fig06}(g) at $\theta\le 30^{\circ}$], and about two orders of magnitude for 
$\varepsilon_b\le 0.4\,{\rm MeV}$  [Fig.\ref{fig06}(c)-Fig.\ref{fig06}(e) and Fig.\ref{fig06}(h)-Fig.\ref{fig06}(j) at $\theta\le 30^{\circ}$]. 
This is another indication of a strong dependence of the Coulomb breakup on the binding energy than the nuclear breakup. 
We further notice in Fig.\ref{fig06}(d,e) that, for $\varepsilon_b\le 0.08\,{\rm MeV}$, the nuclear breakup cross section becomes slightly 
dominant over the total breakup cross section at backward angles ($\theta\ge 90^{\circ}$), while in Fig.\ref{fig06} (left panel), both total 
and nuclear breakup cross sections are hardly distinguishable at $\theta\ge 60^{\circ}$.\\
\begin{figure}[t]
\begin{center}
  \resizebox{82mm}{!}{\includegraphics{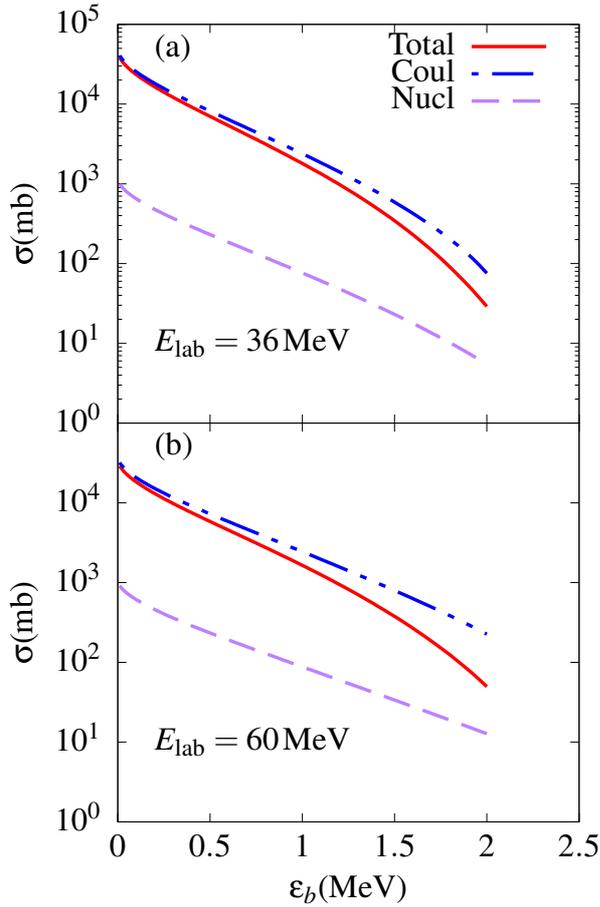}}
\end{center}
\caption{\label{fig08}$^8{\rm Li}+{}^{208}{\rm Pb}$ angular-integrated total, Coulomb and nuclear breakup cross sections as functions 
of the binding energy.}
\end{figure}
\begin{figure}[t]
\begin{center}
  \resizebox{82mm}{!}{\includegraphics{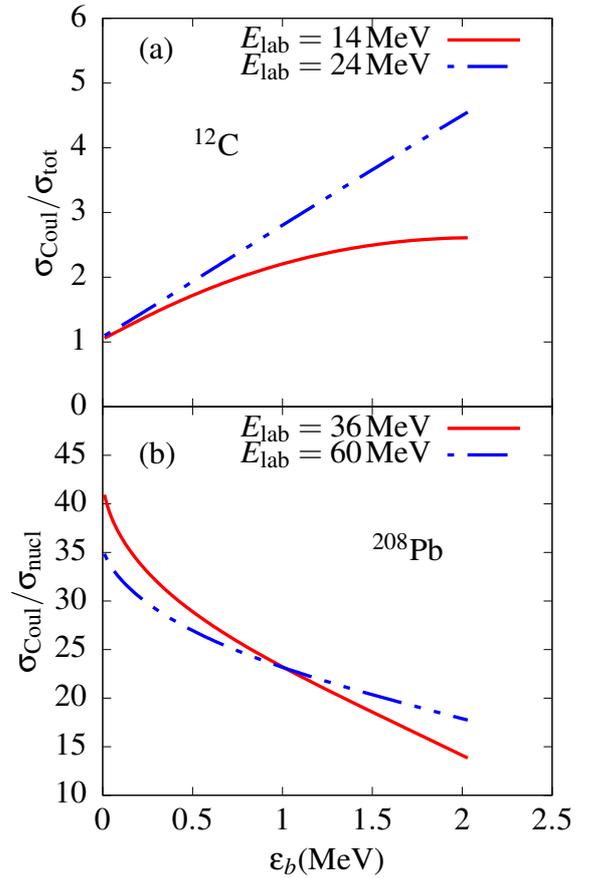}}
\end{center}
\caption{\label{fig09}Ratios $\sigma_{\rm Coul}/\sigma_{\rm nucl}$ and $\sigma_{\rm Coul}/\sigma_{\rm nucl}$ as 
functions of the ground-state binding energy for both targets.}
\end{figure}
\indent As pointed out in the introduction, from a dynamics point of view, due to the fact that the breakup process becomes more peripheral 
as the binding energy decreases, one would expect a decrease of the nuclear breakup cross section, owing to the short-range nature of the 
nuclear forces. However, since it is the opposite that is rather observed, this can be mainly attributed to the static effect, which is associated 
with a longer tail of the ground-state wave function.  This wave function extends more to peripheral regions as the binding energy decreases, 
enabling nuclear forces to exhibit a fair effect in these regions. Therefore, one can argue that, the dependence of the nuclear breakup on the 
binding energy emanates mainly from the static effect, whereas for the Coulomb breakup, this dependence comes from both dynamic and 
static effects, the latter effect being justified by Fig.~\ref{fig01}. This can be regarded as the main reason why the Coulomb breakup is more 
dependent on the projectile ground-state binding energy than the nuclear breakup, as already anticipated. The increase of both Coulomb 
and nuclear breakup cross sections as the binding energy decreases, points to its strong effect on the Coulomb-nuclear interference on 
this energy. It follows that this dependence emanates from both dynamic and static effects. 
\\ 
\begin{figure*}[t]
\begin{center}
  \resizebox{170mm}{!}{\includegraphics{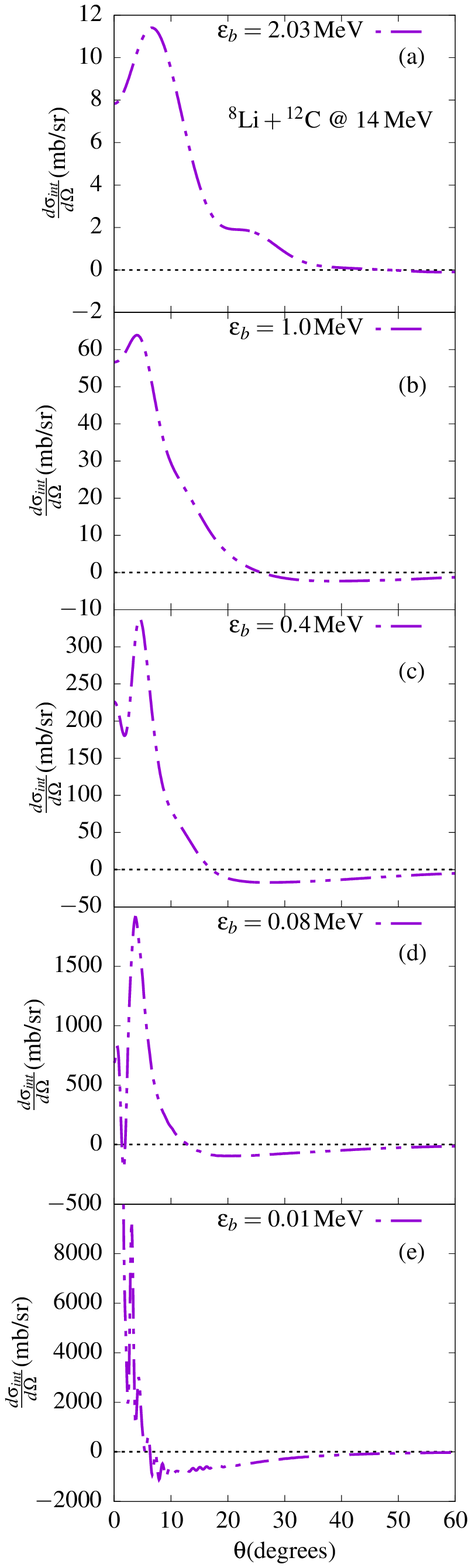}\includegraphics{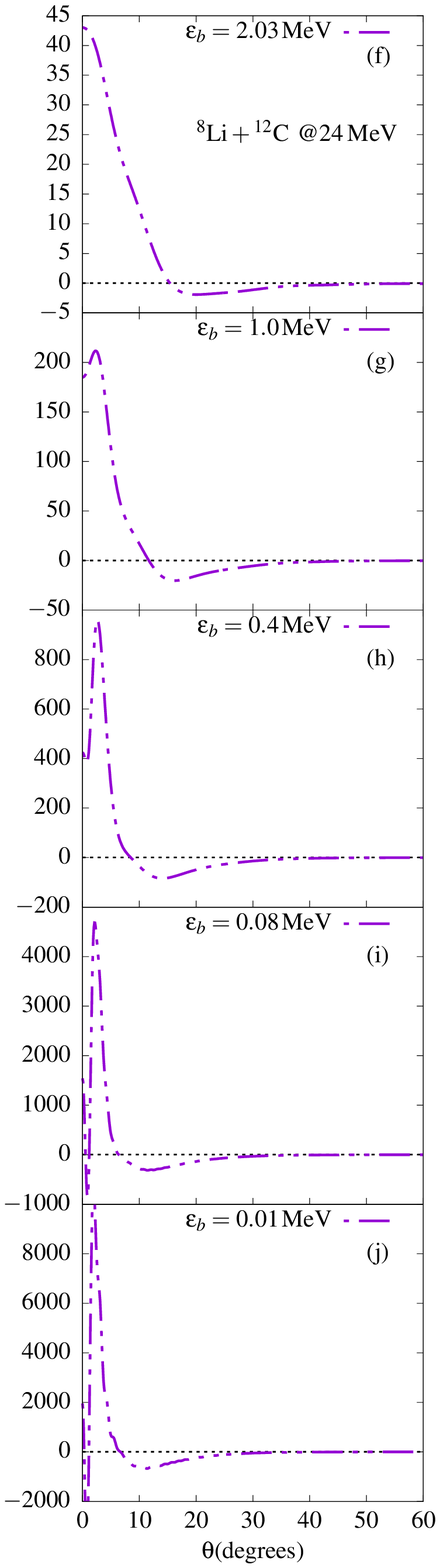}\includegraphics{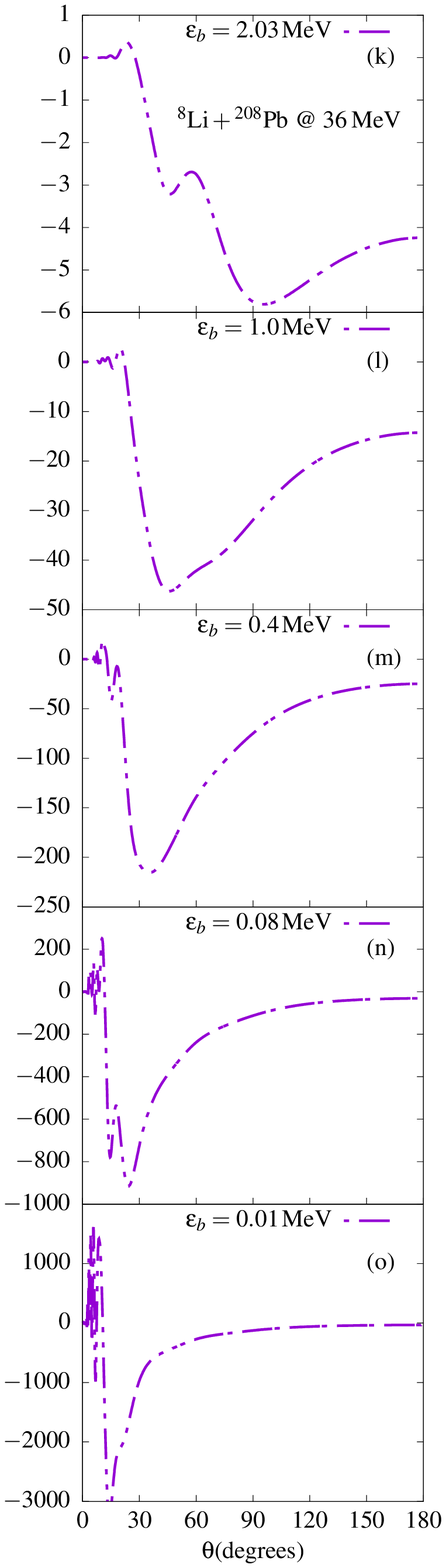}}
\end{center}
\caption{\label{fig10}
Coulomb-nuclear interference for different projectile binding energies (as indicated inside the panels), given by the corresponding angular
differential distributions. 
Left-set of panels are for $^{12}$C at 14 MeV; center-set for $^{12}$C at 24 MeV;
 and right-set for $^{208}$Pb at 36 MeV.
}
\end{figure*}
\indent For a quantitative analysis of these results, we consider the integrated total ($\sigma_{\rm tot}$), 
Coulomb ($\sigma_{\rm Coul}$), and nuclear ($\sigma_{\rm nucl}$) breakup cross sections, which are displayed 
in Fig.~\ref{fig07} for $^{12}{\rm C}$ target, and in Fig.~\ref{fig08} for $^{208}{\rm Pb}$ target. As we have already noticed, 
one clearly sees in both figures that, for $\varepsilon_b\to 0.01\,{\rm MeV}$, $\sigma_{\rm Coul}\gg \sigma_{\rm nucl}$, 
whereas $\sigma_{\rm Coul}< \sigma_{\rm nucl}$ for $\varepsilon_b\ge 0.5\,{\rm MeV}$, [Fig.~\ref{fig07}(a)] and 
$\varepsilon_b> 0.01\,{\rm MeV}$  [Fig.~\ref{fig07}(b)]. One further observes in Fig.~\ref{fig07}(a) that, for 
$\varepsilon_b\to 0.01\,{\rm MeV}$, $\sigma_{\rm Coul}\to \sigma_{\rm tot}$ while for $\varepsilon_b\to 2.03\,{\rm MeV}$, 
$\sigma_{\rm nucl}\to \sigma_{\rm tot}$. In Fig.\ref{fig08}, on the other hand, it is also noticed that, for 
$\varepsilon_b\to 0.01\,{\rm MeV}$, $\sigma_{\rm Coul}\to \sigma_{\rm tot}$, whereas for 
$\varepsilon_b\to 2.03\,{\rm MeV}$, $\sigma_{\rm Coul}> \sigma_{\rm tot}$. We continue the analysis 
considering the $\sigma_{\rm Coul}/\sigma_{\rm tot}$ and $\sigma_{\rm Coul}/\sigma_{\rm nucl}$ ratios. 
It is deduced in Fig.\ref{fig07}(a), that for $\varepsilon_b=2.03\,{\rm MeV}$, the ratio $\sigma_{\rm Coul}/\sigma_{\rm tot}$ 
is about 0.06, and about 0.98 for $\varepsilon_b=0.01\,{\rm MeV}$. At $E_{ lab}=24\,{\rm MeV}$ [Fig.\ref{fig07}(b)], this 
ratio is about 0.12, for $\varepsilon_b=2.03\,{\rm MeV}$, and about 0.65 for $\varepsilon_b=0.01\,{\rm MeV}$. For $^{208}{\rm Pb}$ 
target [Fig.\ref{fig08}(a)], the ratio $\sigma_{\rm Coul}/\sigma_{\rm tot}$ is found to be about 2.61 for $\varepsilon_b=2.03\,{\rm MeV}$, 
and  about 1.06 for $\varepsilon_b=0.01\,{\rm MeV}$. It is about 4.55 for $\varepsilon_b=2.03\,{\rm MeV}$, and about 1.1, for 
$\varepsilon_b=0.01\,{\rm MeV}$, at $E_{ lab}=60\,{\rm MeV}$. Regarding the ratio $\sigma_{\rm Coul}/\sigma_{\rm nucl}$, 
for $^{12}{\rm C} (E_{ lab}=14\,{\rm MeV})$, about 0.08, for $\varepsilon_b=2.03\,{\rm MeV}$, and about 2, for 
$\varepsilon_b=0.01\,{\rm MeV}$.  For $^{208}{\rm Pb}$ target $(E_{ lab}=36\,{\rm MeV})$, it is about 14 for 
$\varepsilon_b=2.03\,{\rm MeV}$, and about 41 for $\varepsilon_b=0.01\,{\rm MeV}$. The trend is similar for the other 
incident energies. These conclusions are nicely portrayed in Fig.\ref{fig09}, where these ratios are plotted as functions 
of the binding energies for both targets. These results highlight the large quantitative importance of the Coulomb breakup
 cross section over the nuclear breakup cross section as the binding energy decreases, even for a light target, owing to
  both dynamic and static breakup effects.

\subsection{Coulomb-nuclear interference}
\label{interf}
\begin{figure}[t]
\begin{center}
  \resizebox{82mm}{!}{\includegraphics{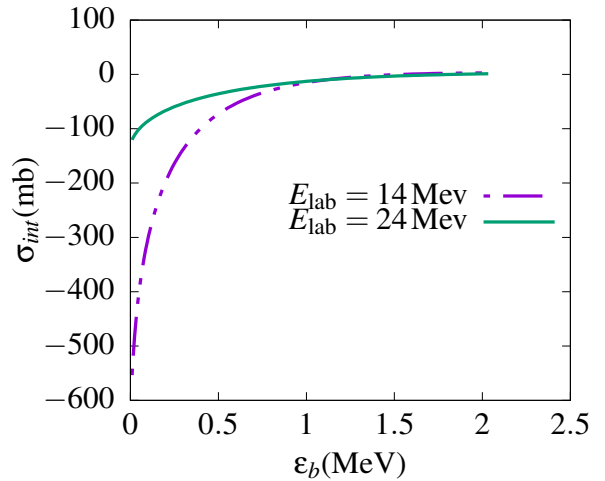}}
\end{center}
\caption{\label{fig11}$^8{\rm Li}+{}^{12}{\rm C}$ integrated Coulomb-nuclear interference as function of the binding energy.}
\end{figure}

\begin{figure}[t]
\begin{center}
  \resizebox{82mm}{!}{\includegraphics{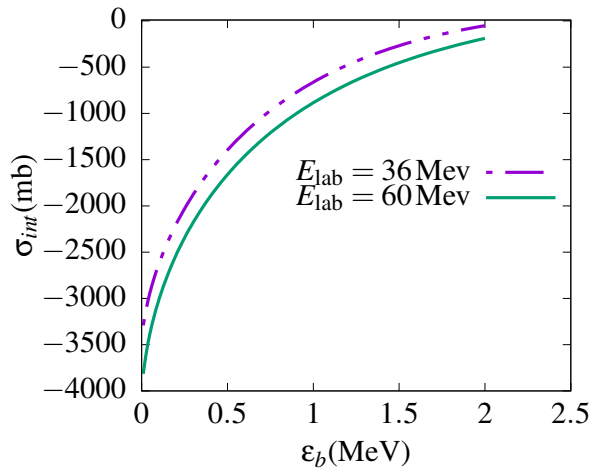}}
\end{center}
\caption{\label{fig12}$^8{\rm Li}+{}^{208}{\rm Pb}$ integrated Coulomb-nuclear interference as function of the binding energy.}
\end{figure}

The analysis performed for different breakup cross sections as functions of the projectile ground-state binding energy
has already paved the way for a better understanding of the dependence of the Coulomb-nuclear interference on this 
energy.
Before considering the details, we first define this interference as given by
\begin{eqnarray}
  \sigma_{\rm int}=\sigma_{\rm tot}-(\sigma_{\rm Coul}+\sigma_{\rm nucl}).
\end{eqnarray}
In other words, we consider the Coulomb-nuclear interference to be the difference between the coherent sum ($\sigma_{\rm tot}$) 
and the incoherent sum ($\sigma_{\rm Coul}+\sigma_{\rm nucl}$) of the Coulomb and nuclear breakup cross sections. 
For the same set of $^8$Li ground-state binding energies, $\varepsilon_b=$0.01 to 2.03 MeV, the differential angular distributions 
$d\sigma_{int}/d\Omega$ are plotted in Fig.~\ref{fig10}, by considering the $^{12}{\rm C}$ target 
at $E_{ lab}=$ 14 MeV (left set of panels) and 24 MeV (center set of panels); and, for the $^{208}{\rm Pb}$ target at 
$E_{ lab}=$36 MeV (right set of panels).
Due to the qualitative similarities of the corresponding results for both 36 and 60 MeV incident energies, for  the 
$^{208}{\rm Pb}$ target we show only one case. 
As already anticipated, it is seen in the panels of this figure that the Coulomb-nuclear interference is strongly strengthened, due  
to the decrease of the binding energy, for both targets, as also concluded in Ref.~\cite{Mukeru15}. By looking at both left and middle set of
panels, one can observe that the interference is strongly constructive at forward angles. The transition angles for destructive interference  
are diminishing as the corresponding values of $\varepsilon_b$ decrease, such that this angle is reduced to around $5^{\circ}$ for
$\varepsilon_b=0.01\,{\rm MeV}$, as shown in the panel (e) of Fig.~\ref{fig10}. 
This indicates that quantitatively, the Coulomb-nuclear interference becomes more destructive as the binding energy decreases. 
It is exclusively destructive for $^{208}{\rm Pb}$ target (Fig.\ref{fig10} right panel), where the destructiveness is strongly 
enhanced, as the binding energy decreases. Therefore, it is clear that the Coulomb-nuclear interference has a strong dependence 
on the projectile binding energy, and becomes strongly destructive as this energy decreases regardless the target mass. To better 
understand the dependence of this interference on the ground-state binding energy, we consider the function $\delta_R(\varepsilon_b)$ 
given in Eq.(\ref{potnucl}). The nuclear breakup dynamics require $\delta_R(\varepsilon_b)\to 0$, as $\varepsilon_b$ increases, 
implying that $R_n\to r_0(A_p^{1/3}+A_t^{1/3})$. We notice that in this case, the Coulomb-nuclear interference is rather weak. 
However, as $\varepsilon_b\to 0$, $\delta_R(\varepsilon_b)$ increase and so does $R_n$. It is seen that this leads to a strong 
interference. This suggests that Coulomb and nuclear forces strongly interfere in the peripheral region, making this interference 
a peripheral phenomenon. On the other hand, $\delta_R(\varepsilon_b)$, depends on the ground-state wave function. One may 
then suggest that the Coulomb-nuclear interference has a strong dependence on the projectile ground-state wave function (static 
effect of the breakup process). Therefore, a better understanding of $\delta_R(\varepsilon_b)$, could shed light on the complexity 
of the Coulomb-nuclear interference. To further elaborate on this, we look back at Figs.\ref{fig07}-\ref{fig09}. We noticed 
in Fig.\ref{fig07}(a) that, qualitatively, $\sigma_{\rm tot}-\sigma_{\rm nucl}\to 0$ as $\varepsilon_b\to 2.03\,{\rm MeV}$. In 
this case, we have that $|\sigma_{\rm int}|\to \sigma_{\rm Coul}$, while 
$\sigma_{\rm tot}-\sigma_{\rm Coul}\to 0\Rightarrow |\sigma_{\rm int}|\to \sigma_{\rm nucl}$ as $\varepsilon_b\to 0.01\,{\rm MeV}$.  
In Fig.\ref{fig07}(b), we observed a similar trend even though, for $\varepsilon_b\to 0.01\,{\rm MeV}$, both nuclear 
and Coulomb breakups contribute significantly to this interference. It is also seen in Fig.\ref{fig08}(a,b) that, 
$\sigma_{\rm tot}-\sigma_{\rm Coul}\to 0\Rightarrow |\sigma_{\rm int}|\to \sigma_{\rm nucl}$, as $\varepsilon_b\to 0.01\,{\rm MeV}$. 
We notice an appreciable contribution from both Coulomb and nuclear breakup cross sections for $\varepsilon_b\to 2.03\,{\rm MeV}$.
\\
\indent Notice that one needs to bear in mind that above we have pure approximate observations. 
For a more rigorous analysis, we have that $|\sigma_{\rm int}|\to \sigma_{\rm nucl}+\Delta_{Ct}$, 
as $\varepsilon_b\to 0.01\,{\rm MeV}$, with $\Delta_{Ct}=\sigma_{\rm Coul}-\sigma_{\rm tot}$, 
as indicated in Figs.~\ref{fig11} and \ref{fig12}, where  integrated Coulomb-nuclear interferences are plotted. 
By observing both, Figs~\ref{fig07} and \ref{fig11} for the light target, one can easily deduce that  
$\Delta_{Ct}\to 0$ as $\varepsilon_b\to 0.01\,{\rm MeV}$. 
However, from Figs.~\ref{fig08} and \ref{fig12}, it is clearly noticed that 
$\Delta_{Ct}\gg \sigma_{\rm nucl}\Rightarrow |\sigma_{\rm int}|\gg \sigma_{\rm nucl}$. 
In fact, for $E_{ lab}=60\,{\rm MeV}$ one can easily deduce that $|\sigma_{\rm int}|\simeq 4\times \sigma_{\rm nucl}$. 
This emphasizes the fact that, for low binding energies, small nuclear breakup contribution does not necessarily amount 
to small Coulomb-nuclear interference, which is the well-known ``Coulomb-nuclear interference problem''. This further 
indicates that this interference has a strong dependence on the function $\delta_R(\varepsilon_b)$.

\section{Conclusion}
\label{conclusion}

In this paper, we have studied the $^8{\rm Li}+{}^{12}{\rm C}$ and $^8{\rm Li}+{}^{208}{\rm Pb}$ breakup reactions, 
in order to investigate 
the dependence of the Coulomb-nuclear interference on the $^8{\rm Li}$ ground-state binding energy $\varepsilon_b$. 
To this end, we select an arbitrary range of $\varepsilon_b$, from 0.01 MeV to 2.03 MeV. 
In the first step, we have checked the effect of these energies on the Coulomb and nuclear breakup cross sections,
when it was found that  
the ground-state binding energy has a stronger effect on the Coulomb breakup cross section than on the nuclear breakup cross section.
This behavior results from 
the fact that, in view of the long-range nature of the Coulomb forces and the electromagnetic transition matrix elements, 
the Coulomb breakup depends more strongly  
on both dynamics and static breakup effects.
In particular, we found that, even for the known nuclear-dominated $^8{\rm Li}+{}^{12}{\rm C}$ reaction, 
while $\sigma_{\rm nucl}\to \sigma_{\rm tot}$ as $\varepsilon_b\to 2.03\,{\rm MeV}$, 
we have $\sigma_{\rm Coul}\to \sigma_{\rm tot}$ as $\varepsilon_b\to 0.01\,{\rm MeV}$, with $\sigma_{\rm nucl}\ll\sigma_{\rm Coul}$. 
The analysis of the Coulomb-nuclear interference as function of the binding energy showed that, quantitatively and 
particularly for a heavy target, it can be written as $|\sigma_{\rm int}|=\sigma_{\rm nucl}+\Delta_{Ct}$, where 
$\Delta_{Ct}=\sigma_{\rm Coul}-\sigma_{\rm nucl}$. It is obtained that, as $\varepsilon_b\to 0.01\,{\rm MeV}$, 
$\Delta_{Ct}\gg \sigma_{\rm nucl}$, 
implying that $\sigma_{\rm int}\gg \sigma_{\rm nucl}$. In fact, for $^8{\rm Li}+{}^{208}{\rm Pb}$ at $E_{ lab}=60\,{\rm MeV}$, 
we deduced that $|\sigma_{\rm int}|\simeq 4\times \sigma_{\rm nucl}$, whereas $\sigma_{\rm Coul}\simeq 35\times \sigma_{\rm nucl}$. 
This is a clear demonstration that a small nuclear contribution in a Coulomb-dominated reaction does not imply insignificant 
Coulomb-nuclear interference. 
We attempted to interpret this through the ``peripherality'' of the nuclear breakup. The range of nuclear forces 
is defined as $R_n=r_0\left(A_p^{1/3}+A_t^{1/3}\right)+\delta_R(\varepsilon_b)$, where $\delta_R(\varepsilon_b)$ is a function 
that depends on the density of the ground-state wave function through the binding energy. 
This function is such that $\delta_R(\varepsilon_b)\to 0$ as $\varepsilon_b$ decreases, and $\delta_R(\varepsilon_b)>0$ as 
$\varepsilon_b$ increases. 
The main conclusion of the present work, obtained quantitatively by the breakup analysis of the effect of projectile 
ground-state energies in light and heavy targets, as verified by $\delta_R(\varepsilon_b)$, is that the Coulomb-nuclear 
interference is a peripheral phenomenon, with the Coulomb and nuclear forces strongly interfering destructively in the peripheral region.

\section*{ACKNOWLEDGEMENTS}
LT is thankful to the following agencies for partial support: 
Conselho Nacional de Desenvolvimento Cient\'\i fico e Tecnol\'ogico [  
INCT-FNA Proc.464898/2014-5 (LT and JL), Proc. 306191-2014-8(LT)], \\
Funda\c c\~ao de Amparo \`a Pesquisa do Estado de S\~ao Paulo [Projs. 2017/05660-0(LT)].

\end{document}